\documentclass{ws-ijmpcs}

\begin{document}

\markboth{D. Deb et al.}
{METHODS TO COMPUTE PRESSURE AND WALL TENSION IN FLUIDS}

%
\catchline{}{}{}{}{}
%

\title{METHODS TO COMPUTE PRESSURE AND WALL TENSION IN FLUIDS CONTAINING HARD
PARTICLES}

\author{DEBABRATA DEB, DOROTHEA WILMS, ALEXANDER WINKLER, \\PETER VIRNAU, and KURT BINDER}

\address{Institut f\"ur Physik, Johannes Gutenberg Universit\"at Mainz\\
Staudinger Weg 7, D 55099 Mainz, Germany\\
deb@uni-mainz.de}

\maketitle

\begin{history}
\received{Day Month Year}
\revised{Day Month Year}
\end{history}

\begin{abstract}
Colloidal systems are often modelled as fluids of hard particles
(possibly with an additional soft attraction, e.g. caused by polymers also
contained in the suspension). In simulations of such systems, the virial
theorem cannot be straightforwardly applied to obtain the components of the
pressure tensor. In systems confined by walls, it is hence also not straightforward
to extract the excess energy due to the wall (the ``wall tension'') from
the pressure tensor anisotropy. A comparative evaluation of several methods
to circumvent this problem is presented, using as examples fluids of hard
spheres and the Asakura-Oosawa model of colloid-polymer mixtures with a size
ratio $q=0.15$ (for which the effect of the polymers can be integrated out to
yield an effective attractive potential between the colloids). Factors
limiting the accuracy of the various methods are carefully discussed, and
controlling these factors very good mutual agreement between the various
methods is found.

\keywords{hard spheres; Asakura-Oosawa model; ensemble-mixing method; Monte Carlo simulation}
\end{abstract}

\ccode{PACS numbers: 82.70.Dd, 05.10.Ln, 61.20.Ja, 68.08.-p}

\section{Introduction}	

In many colloidal dispersions, the interaction between the colloidal particles
(or at least the repulsive part of these interactions) can be represented as that of
hard particles (hard spheres, hard rods, etc.) \cite{1,2,3,4,5}. Colloidal dispersions
are model systems for the experimental study of cooperative phenomena in condensed
matter, since the size of these particles (in the $\mu m$ range) and their slow
dynamics allows their study in unprecedented detail, which would not be possible for condensed
matter systems formed from small molecules. Moreover, hard particles are the
archetypical many-body systems, for which computer simulation methods were first
introduced a long time ago: the importance sampling Monte Carlo method \cite{6}
was introduced discussing a fluid of hard disks; similarly, Molecular
Dynamic methods were introduced in the context of the study of the hard sphere
fluid \cite{7}. These fluids formed from hard particles also are an important
testing ground for analytical theories of fluids \cite{8}. The above remarks
just sketch a few motivations to explain why the simulation of fluids
formed from hard particles is of fundamental importance.

However, simulations of fluids containing hard particles suffer from one important
technical difficulty: the well-known virial equation for the pressure tensor
(for systems where particles interact with pair-wise forces $\vec{f} (\vec{r}_{ij})$
with $r_{ij}=| \vec{r}_{ij} |, \vec{r}_{ij} =\vec{r}_i-\vec{r}_j)$ \cite{8}

\begin{equation} \label{eq1}
p^{\alpha \beta} = \rho k_BT \, \delta^{\alpha \beta} + \frac{1}{dV} \Big \langle
\sum\limits_{i < j} f^\alpha (\vec{r}_{ij}) r^\beta_{ij} \Big \rangle _{NVT}
\end{equation}

is not straightforwardly applicable (for hard spheres of diameters $\sigma_i$,
the force $\vec{f} \equiv 0$ if $r_{ij} > (\sigma_i + \sigma_j)/2$, while
distances $r_{ij} < (\sigma_i + \sigma_j)/2$ are excluded, and for
$r_{ij}=(\sigma_i + \sigma_j)/2$ the potential energy jumps from zero to
infinity, and hence the force is also not defined in this case). Note that in
Eq.~(\ref{eq1}) $\alpha, \beta$ denote Cartesian components, $\rho=N/V$ is the
density ($N$=particle number, $V$= ``volume'' of the $d$-dimensional system),
$k_B$ is Boltzmann's constant, $T$ absolute temperature, and $\langle \cdots \rangle_{NVT}$
is understood as a statistical average in the canonical ($NVT)$ ensemble.
Computation of the pressure is not only of the interest to clarify the
equation of state of the considered system in the bulk, but is
also required when one considers a system confined by walls, and wishes to
estimate the excess free energy due to the walls. Assuming a geometry where
the system is confined by two identical walls at a (large) distance $D$ apart, the wall-fluid surface excess free energy per unit area $\gamma_{wf}$ (also called ``wall tension'') is
expressed in terms of the anisotropy of the pressure tensor as \cite{9}

\begin{equation} \label{eq2}
\gamma_{wf}=D(P_N-P_T)/2
\end{equation}

where $P_N=p^{zz}$ is the ``normal pressure'' which in equilibrium is
homogeneous in the system, while $P_T$ is the average tangential pressure

\begin{equation} \label{eq3}
P_T= (1/2D)\int\limits_0^D (p^{xx} (z) + p^{yy} (z)) dz .
\end{equation}

Note that Eq.~(\ref{eq2}) supposes that the two walls are far enough from
each other that in between the fluid can attain its bulk properties,
i.e. the two walls are non-interacting, the density profile $\rho(z)$ as well
as the transverse pressure tensor components $p^{xx}(z)$ and $p^{yy}(z)$ are constant,
independent of $z$, for a broad region of $z$ near $z=D/2$. In contrast,
$\rho(z)$ does depend on $z$ near the walls, as well as $p^{xx}(z)$ and
$p^{yy} (z)$, while the transverse component $p^{zz}(z)$ is strictly
independent of $z$, irrespective of the choice of $D$. The generalization
of $p^{\alpha \beta}$ in Eq.~\ref{eq1}, which addresses only the behavior
in the bulk, to such inhomogeneous systems with external boundaries will be
discussed in Sec. 3 below.

In the present paper, we shall discuss methods from which the pressure and
the wall tension can be computed for fluids where particles experience
hard-core interactions, and hence the straightforward use of Eq.~\ref{eq1},
or its extension to inhomogeneous systems is not possible. In Sec.~2, we shall describe
the application of a recent method due to Schilling and Schmid \cite{10}, to
compute absolute free energies of disordered structures by molecular simulations,
considering the situation described above, where a system is confined by walls.
We assume planar, structureless walls, with a smooth repulsion described by a
Weeks-Chandler-Andersen (WCA) \cite{11}-type potential,

\begin{equation} \label{eq4}
V_{\rm WCA}(z) = 4 \varepsilon[(\sigma_w /z)^{12} - (\sigma_w/z)^6 + 1/4], \,
0 \leq z \leq z_c= \sigma_w 2^{1/6} \quad ,
\end{equation}

while $V_{WCA} (z \geq z_c)=0$. Here the strength ($\varepsilon$) and
range $(\sigma_w)$ of the potential provide parameters which can be
varied, and so also the wall tension $\gamma_{wf}$ can be changed. For the
chosen geometry, the free energy density can be written as

\begin{equation} \label{eq5}
f(\rho, T, D) = f_b (\rho, T) + (2/D) \gamma_{wf}
\end{equation}

where $f_b( \rho, T)$ is the free energy density of a corresponding
bulk system at density $\rho$ and temperature $T$. We shall test
the applicability of Eq.~(\ref{eq5}) for the simplest case,
namely the hard
sphere fluid.

In Sec.~3 we follow the alternative approach of De Miguel and Jackson
\cite{12}, where Eqs.~(\ref{eq2}),~(\ref{eq3}) are used for hard
particles in spite of the fact that a straightforward use of Eq.~(\ref{eq1})
is not possible. In this method, virtual changes of simulation box linear
dimensions are performed to sample the probability that no molecular pair overlaps would
occur as a result of these virtual moves. From this probability, the pressure
can be estimated. While we already have verified the accuracy of this method 
for the case of a hard sphere fluid \cite{13}, we here extend the method
to the case where the interparticle potential has both a hard core
repulsion and a soft attraction. A well-known and practically relevant
example for such a situation is provided by the Asakura-Oosawa model
\cite{14,15,16} for colloid-polymer mixtures, in the limit where the
polymers are much smaller than the colloids. Describing the colloids as
hard spheres of diameter $\sigma$ and the polymers as soft spheres of
diameter $\sigma_p$, polymer-colloid overlap is also strictly forbidden,
but polymer coils can interpenetrate and hence overlap with negligible energy
cost. For $q \equiv \sigma_p/\sigma \leq 0.154$, one can integrate out
the polymer degrees of freedom from the partition function exactly,
replacing their physical effects by an effective attractive interaction
between the colloids \cite{14,15,16,17,18}

\begin{equation} \label{eq6}
V_{\rm AO} (r) / k_BT = - \eta ^r_p \Big(\frac{1 +q}{q}\Big)^3 \Big\{1 - \frac{3 r/\sigma}{2 (1+q)} +
\frac{(r/\sigma)^3}{2 (1 + q)^3} \Big\}, \quad \sigma < r < \sigma + \sigma _p,
\end{equation}

while $V_{\rm AO} (r < \sigma)= \infty$, $V_{AO} (r \geq \sigma + \sigma_p)=0$. The strength
of this attractive potential is controlled by the ``polymer reservoir packing fraction''
$\eta_p^r$, which is related to the chemical potential $\mu_p$ of the polymers
as $\eta^r_p= (\pi / 6) \sigma^3_p \exp (\mu_p/k_BT)$.

Both methods of Secs 2. and 3 are checked by applying an independent third method,
extending \cite{13} a thermodynamic integration method \cite{19} where one
gradually inserts a wall potential in a system without walls (with periodic boundary conditions).
Our extension \cite{13} called ``ensemble mixing method`` samples the free energy of
systems with intermediate Hamiltonians. This new method is presented in Sec.~4, while Sec.~5
presents our conclusions.

\section{Estimation of wall tension from computation of absolute free energies}

As is well known, the absolute free energy of a system is normally not a straightforward output
of a Monte Carlo simulation \cite{20}. The most commonly used strategy is to use
thermodynamic integration, supposing that the Hamiltonian $\mathcal{H}(\lambda)$ depends
on a parameter $\lambda$ that can be varied from a reference state (characterized by $\lambda_0$)
whose free energy is known, to the state of interest ($\lambda_{1}$), without crossing any
phase transition,

\begin{equation} \label{eq7}
\Delta F = F (\lambda_1) - F ( \lambda_0)=\int\limits_{\lambda_0}^{\lambda_1} d \lambda' \langle
\mathcal{H} (\lambda') / \partial \lambda' \rangle_{\lambda'} \quad .
\end{equation}

For a dense fluid, Schilling and Schmid \cite{10} propose to take as a reference state a configuration
that is representative for the structure of interest (obtained within a standard simulation of the
considered system). From this well-equilibrated state a reference state is constructed by fixing this
particular configuration with suitable external potentials that hold all particles rigidly in their
position $\{\vec{r}^{\;p}_i\}$ in that particular configuration (and the internal interactions between
the particles can be switched off). In practice, Schilling and Schmid \cite{10} used as a pinning
potential in the reference state

\begin{equation} \label{eq8}
U_{\rm ref} (\lambda) = \lambda \sum_i \phi (|\vec{r}_i-\vec{r}_i^{\;p}|/r_{\rm cut}), \,\,
{\rm with} \, \phi(x)=x-1\, ,
\end{equation}

 where it is understood that particle $i$ can only be pinned by well $i$ at $\vec{r}_i^{\;p}$, and
 not by the other wells (but one carries out identity swaps to make the particles indistinguishable).
For this non-interacting reference state the (configurational) reference free energy is

\begin{equation} \label{eq9}
F_{\rm ref}(\lambda)=\ln (N/V) - \ln [1+(V_0/V) g_\phi(\lambda)] \quad ,
\end{equation}

 where $V_0=4 \pi r^3_{\rm cut}/3$ and $g_\phi(\lambda)=3 \lambda^d (e^\lambda - \sum\limits_{k=0}^3
 e^k /k!)$, for the above choice of $\phi(x)$. Then, intermediate models are defined as

 \begin{equation} \label{eq10}
 \mathcal{H}'(\lambda)= \mathcal{H}_{\rm int} + U_{\rm ref} (\lambda) \quad,
 \end{equation}

 where $\mathcal{H}_{\rm int}$ describes the interactions in the system which then are switched on
 (if necessary, in several steps). The free energy difference relative to $F_{\rm ref}(\lambda)$
 then is computed by thermodynamic integration, for which $\langle \partial \mathcal {H}_{\rm ref}
 (\lambda)/\partial \lambda \rangle =\langle \sum_i \phi(|\vec{r}_i - \vec{r}_i^{\;p} |/r_{\rm cut}) \rangle_\lambda$ needs
 to be sampled. For details, how this is done efficiently, we refer to the original publication \cite{10}.
 Schilling and Schmid \cite{10} have already tested this method for hard spheres in the bulk, obtaining
 the free energy for several densities, and verifying the (expected) agreement with the Carnahan-Starling
 equation of state \cite{21}.

\begin{figure}[pb]
\centerline{\psfig{file=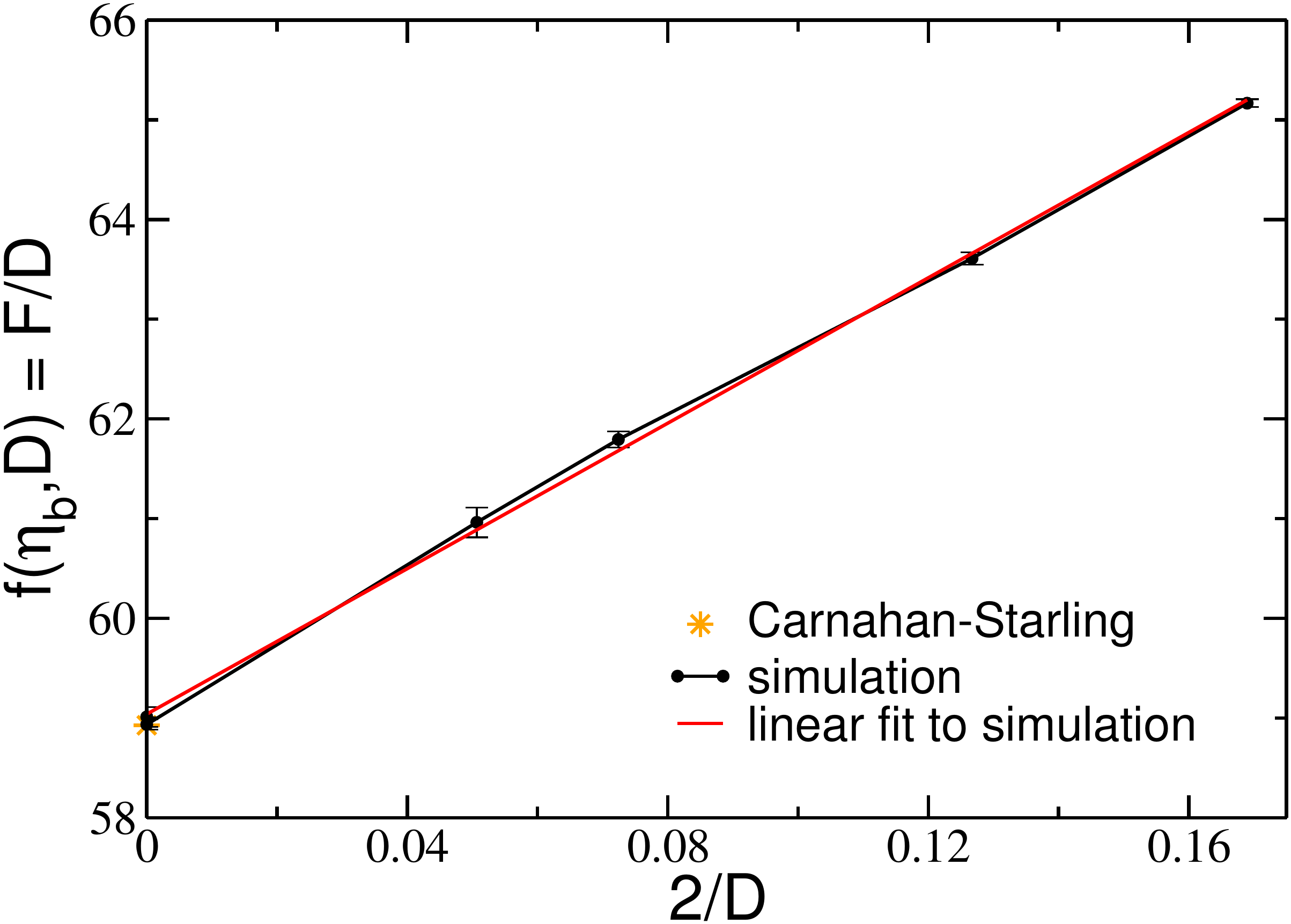,width=6.7cm}}
\vspace*{8pt}
\caption{\label{fig1} Extrapolation of the free energy $f(\eta_c, D)$ of a hard sphere
system at a packing fraction $\eta_c=0.3686$ of colloids plotted versus $2/D$, $D$ being
the distance between two parallel walls, at which the WCA potential (Eq.~(\ref{eq4})) acts.
Note that the intercept at the ordinate is known accurately from the Carnahan-Starling
equation of state. The slope of the straight line fit yields the wall tension.}


\end{figure}

In the present section, we present a straightforward extension of this method to a hard sphere system
in a $L \times L \times D$ geometry, choosing $L=6$ (lengths being measured in units of $\sigma$), choosing
a packing fraction $\eta=(\pi/6)\rho=0.3686$ and several choices for $D$ (Fig.~\ref{fig1}). In order
to obtain a reliable estimation of the error of this method, the approach was repeated for each
choice of $D$ for 5 independent (but equivalent) reference configurations. The total number of Monte Carlo
steps per particle for each of these runs was in the order of about $4 \cdot 10^5$ for each of the $500$ steps of the thermodynamic integration and about $2\cdot10^6$ for each of the $40$ steps in which the potential wells were switched on. Fig.~\ref{fig1} shows that indeed $F/D$ plotted vs. $1/D$ is compatible with a straight line,
 and the intercept of this straight line agrees with the Carnahan-Starling result (which was verified
 directly by also running a $L \times L \times L$ system at the chosen packing fraction).
 The accuracy with which the straight line can be fitted to the data points allows us to estimate
 $\gamma_{wf}$ from Eq.~(\ref{eq5}) with a relative accuracy of about 1\%, and the result (choosing
 units such that $k_BT=1$) $\gamma=1.01 \pm 0.01$ is fully compatible with the result of the ensemble
 mixing method for this case (Sec. 5). Thus, Fig.~\ref{fig1} is a valuable feasibility test that
 shows that the Schilling-Schmid method \cite{10} is useful for the estimation of interfacial excess
 free energies of fluids.

 \section{Computation of the pressure tensor in systems of hard particles confined by walls}

 In this section we consider the AO-model of colloid-polymer mixtures, where particles interact both
 with a hard core interaction (for $r< \sigma)$ and a soft attractive part of the interaction, as
 described in Eq.~(\ref{eq6}). We wish to apply Eq.~(\ref{eq2}), and need to compute both $P_N$ and
 $P_T$ for this purpose. Furthermore we allow again for walls where the WCA potential, Eq.~(\ref{eq4}),
 acts.

 First we note that in a system which is inhomogeneous in the $z$-direction the generalization of
 Eq.~(\ref{eq1}) to a local pressure tensor can be written as \cite{22}

 \begin{eqnarray} \label{eq11}
 && p^{\alpha \beta}(z) = \rho (z) k_BT \delta^{\alpha \beta} - \frac{1}{L^2} \sum\limits_{i < j}
 r^\alpha_{ij} \Big(\frac{\partial U}{\partial \vec{r}_{ij}} \Big)^\beta \theta[(z-z_i)/z_{ij}]
 \theta[(z_j-z)/z_{ij}]/{|z_{ij}|}\nonumber\\
 && - \rho (z) z \frac{d}{dz} V_{\rm WCA} (z) \delta^{\alpha z} \delta^{\beta z} \quad .
 \end{eqnarray}

 Note that in the total force acting on a particle there is also a contribution due to the wall potential,
 which has been written down explicitly in the last term, which contributes only for $\alpha=z$
 and $\beta=z$. Thus, we can decompose $p^{zz}$ into the three contributions $\rho(z) k_BT$
 and $p_{\rm wall} (z)\equiv- \rho (z)z(d V_{\rm WCA} (z)/dz)$ and the part due to interparticle interaction $p ^{zz}_{\rm int}(z)$,

 \begin{equation} \label{eq12}
 p^{zz}=\rho(z) k_BT + p ^{zz}_{\rm int}
 (z) + p_{\rm wall}(z) \quad ,
 \end{equation}

 noting that while all three components of $p^{zz}$ do have a pronounced $z$-dependence in
 Eq.~(\ref{eq12}), the total normal pressure $p^{zz}$ in fact must be independent of $z$.

 In our case Eqs.~(\ref{eq11}),~(\ref{eq12}) can be applied directly with respect to the soft attractive
 part of the interactions, but not with respect to the hard core part. Here we apply the principle
 that all contributions to the pressure simply are additive,

 \begin{equation} \label{eq13}
 p^{\alpha \beta}_{\rm int} (z) =p^{\alpha \beta}_{\rm int,\, soft}(z) + p^{\alpha \beta}_{\rm int,\, hard} (z) \quad ,
 \end{equation}

 and we only compute $p^{\alpha \beta}_{\rm int,\, soft} (z)$ directly from the virial expression, i.e. the second term on the right hand side of Eq.~(\ref{eq11}), while for $p^{\alpha \beta}_{\rm int, \, hard} (z)$ we apply the method of De Miguel and Jackson \cite{12}, as already used in our previous work on the hard sphere fluid \cite{13}.

 In the latter method, one considers virtual volume changes according to

 \begin{equation} \label{eq14}
 D \rightarrow D'=D (1- \xi), \, L \rightarrow L'=L (1-\xi)
 \end{equation}

 where $\xi<<1$. In any equilibrium configuration, there cannot be any overlaps
 between the hard cores of any pair of particles, but in dense fluids these virtual
 moves will create some overlaps. One then can sample the probabability $P_{\rm nov}(\xi)=
 \exp (-b \xi)\approx 1-b \xi$ that no overlap of particles occurs. Denoting the constant
 $b$ for the moves from $D$ to $D'$ as $b_N$, and for moves from $L$ to $L'$ as $b_T$,
 we obtain

 \begin{equation} \label{eq15}
 P_{N,\, {\rm hard}}/( \rho k_BT)= 1 + b_N/N,
 \end{equation}

 \begin{equation} \label{eq16}
 P_{T, \, {\rm hard}}/(\rho k_BT)=1 + b_T/N \quad ,
 \end{equation}

 where

 \begin{equation} \label{eq17}
 P_{N, {\rm hard}} = (1/D) \int\limits_0^D p^{zz}_{\rm int,\, hard} (z) dz\quad ,
 \end{equation}

   and

   \begin{equation} \label{eq18}
   P_{T,\, {\rm hard}} =(1/2D) \int\limits^D_0 [p^{xx}_{\rm int, \, hard} (z) + p^{yy}_{\rm int, \, hard}(z)]
   dz. \quad .
   \end{equation}

   In view of the substantial difficulty in obtaining very accurate simulation ``data'', we have not
   attempted to compute the individual profiles $p^{\alpha \beta}_{\rm int} (z)$, noting
   that for the desired use of Eq.~(\ref{eq2}) only the quantities averaged across the film matter,
   as written in Eqs.~(\ref{eq17}),~(\ref{eq18}).

\begin{figure}
\begin{center}$
\begin{array}{ccc}
\includegraphics[width=0.5\textwidth]{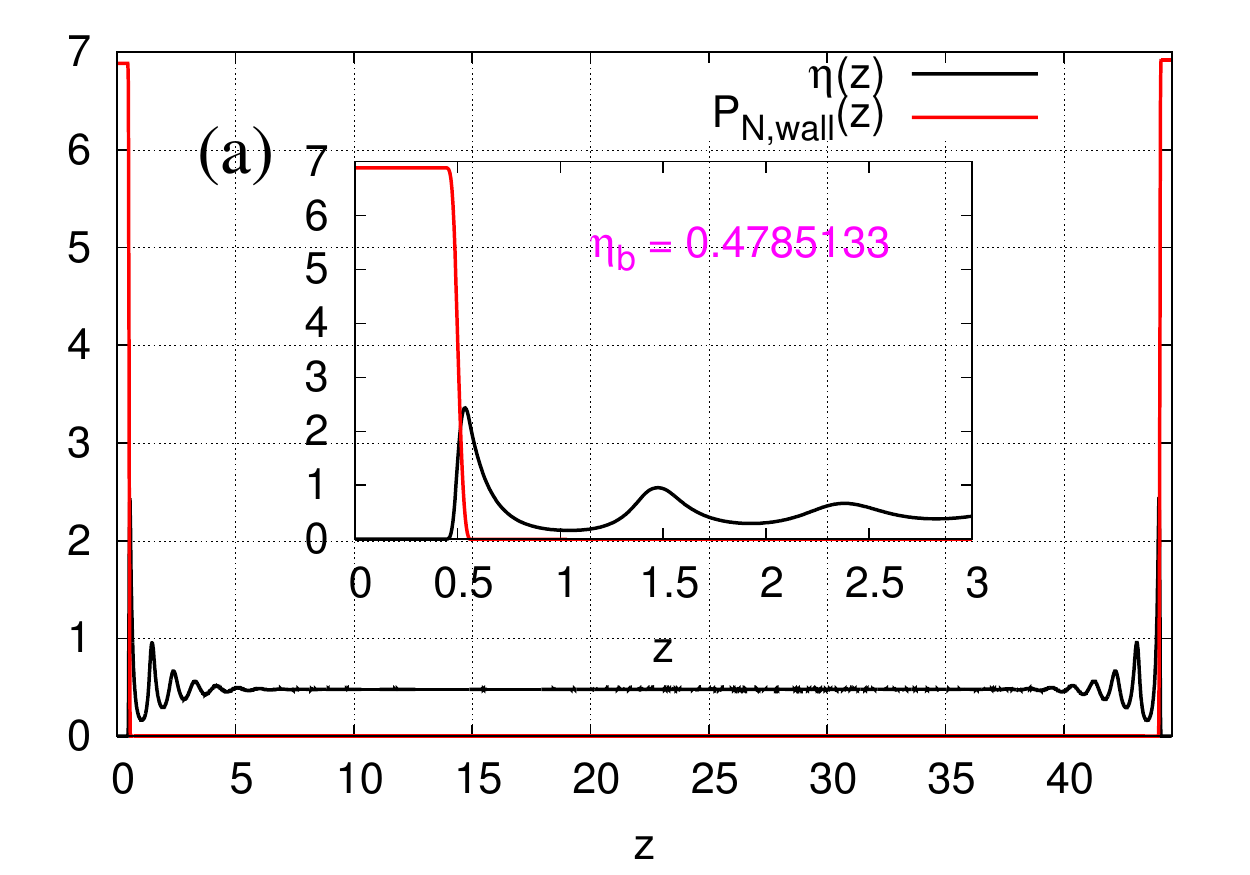} & \hspace*{0.0cm}&
\includegraphics[width=0.5\textwidth]{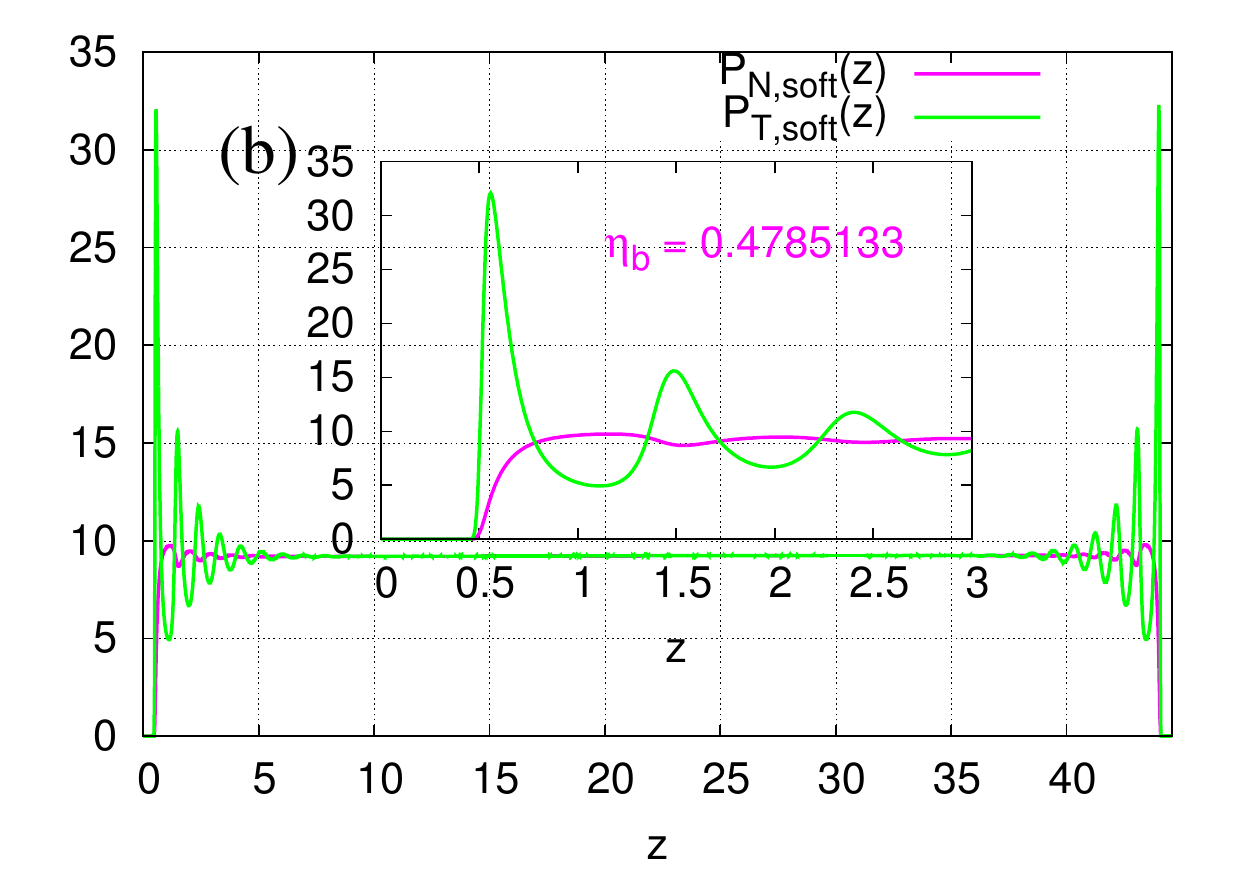}
\end{array}$
\caption{ \label{fig2} Profiles of the packing fraction $\eta(z)$ and wall contribution $p^{zz}_{\rm soft} (z)$
to the normal pressure $p_{\rm wall}(z)$, case (a), and of the attractive potential parts to both
normal $\{p^z_{\rm soft}(z)\}$ and tangential $\{p^{xx}_{\rm soft} (z)= p ^{yy}_{\rm soft}(z)\}$
pressure tensor, for the AO model at bulk packing fraction $\eta_b=0.4785133$. Note that
$p^{xx}_{\rm soft} (z)$, $p^{zz}_{\rm soft}(z)$ are obtained from the second term on the
right hand side of Eq.~(\ref{eq1}). Insets show the variations near the left wall on magnified abscissa scales.}
\end{center}
\end{figure}

As an example, Fig.~\ref{fig2} shows the density profile $\rho(z)$ and the profile $p_{\rm wall}(z)$
as well as the profiles of those parts of both $p^{zz}_{\rm int}(z) $ and $p_{\rm int}^{xx} (z)= p^{yy}_{\rm int}
(z)$ that are due to attractive parts of the potential of the AO model (Eq.~\ref{eq6}). We recognize
that all these contributions exhibit rapid large oscillations near the walls. These oscillations need to
be well resolved (with a binning of the $z$-coordinate that is about an order of magnitude smaller
than the hard sphere diameter, which sets the scale for these oscillations) and carefully sampled. 
Otherwise systematic errors are made when these contributions are integrated over $z$, and despite
of the simplification that only integrated pressures are needed (cf. Eqs.~(\ref{eq17}),~(\ref{eq18}))
all results would suffer from systematic errors.

   However, even with this simplification it is very difficult to obtain a very good accuracy,
   because systematic errors that occur if the range of $\xi$ for which $P_{\rm nov}(\xi)$ is
   sampled are not extremely small.

\begin{figure}
\begin{center}
\includegraphics[width=0.75\textwidth]{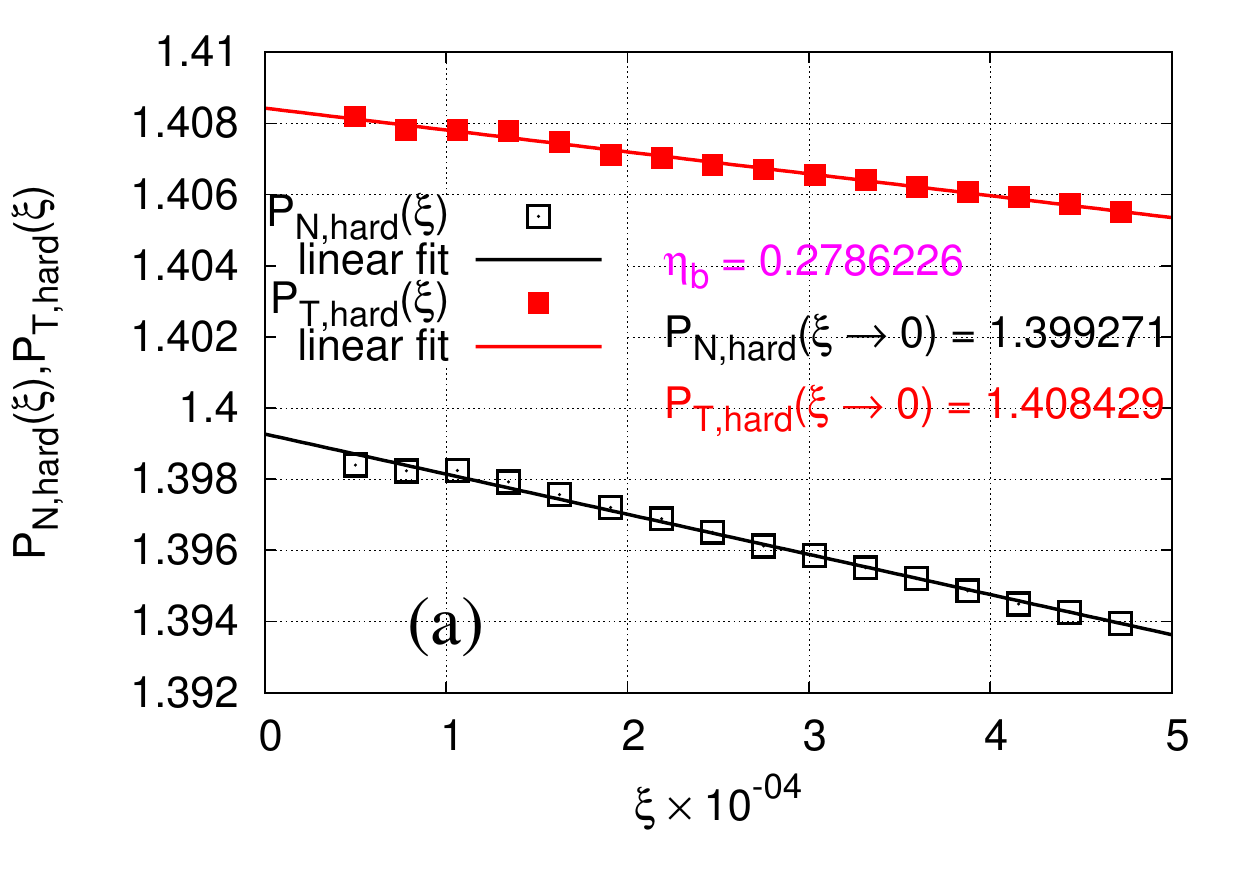}\\
\includegraphics[width=0.75\textwidth]{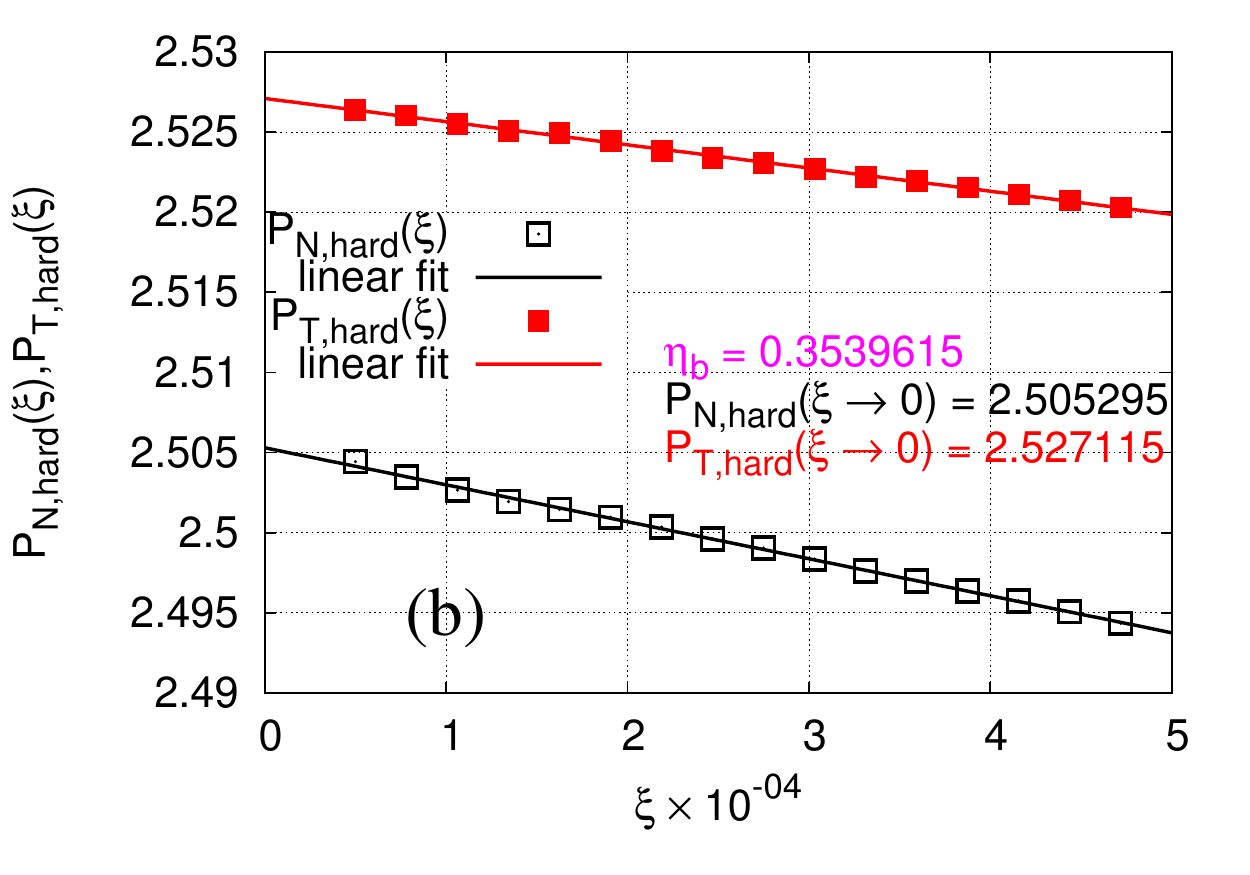}\\
\includegraphics[width=0.75\textwidth]{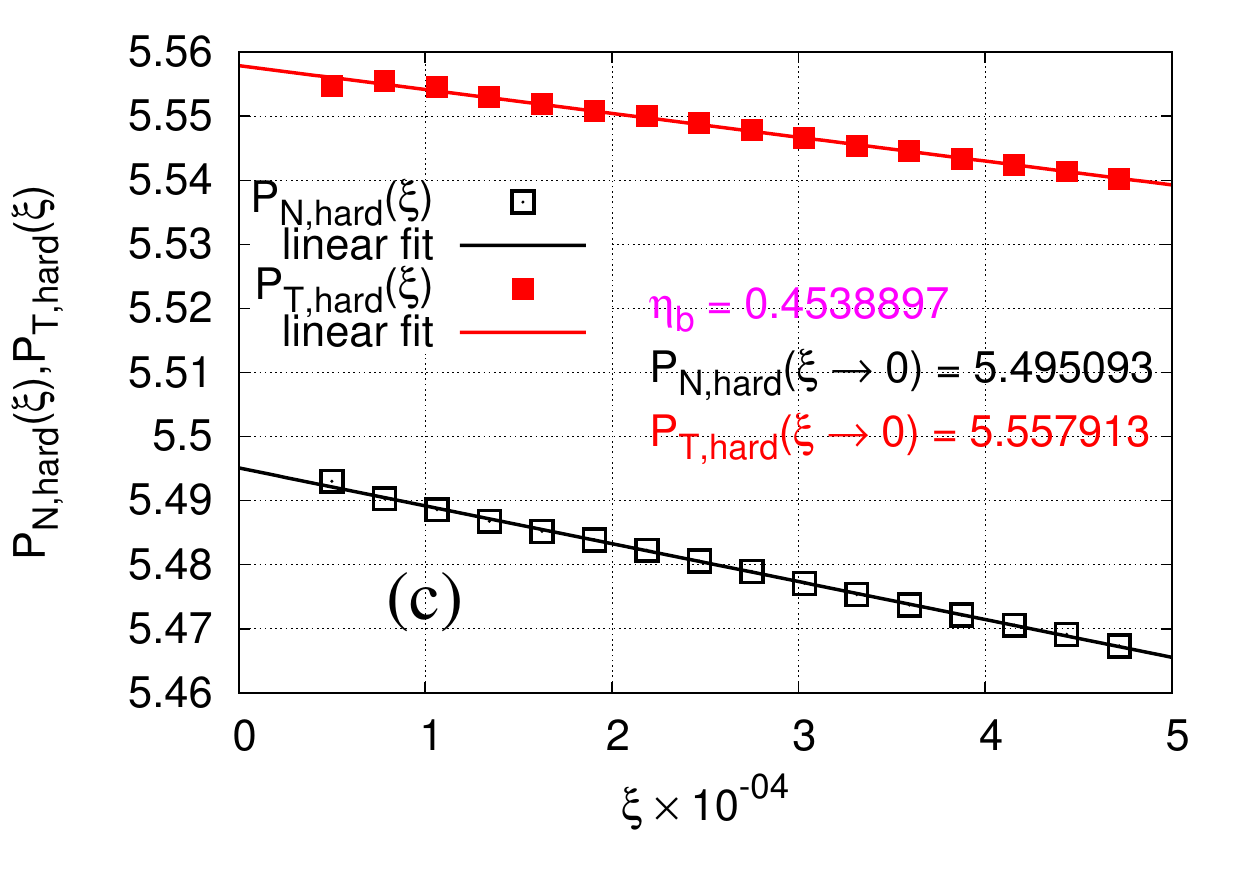}
\caption{\label{fig3} Extrapolation of the repulsive part of the total normal pressure
$P_{N, {\rm hard}}$ and the total tangential pressure $P_{T, {\rm hard}}$ resulting from
the hard sphere interactions at three packing fractions plotted vs $\xi$ for $\eta_b= 0.2786226$(a), $0.3539615$(b) and $0.4538897$(c).}
\end{center}
\end{figure}

\begin{figure}
\begin{center}$
\begin{array}{ccc}
\includegraphics[width=0.5\textwidth]{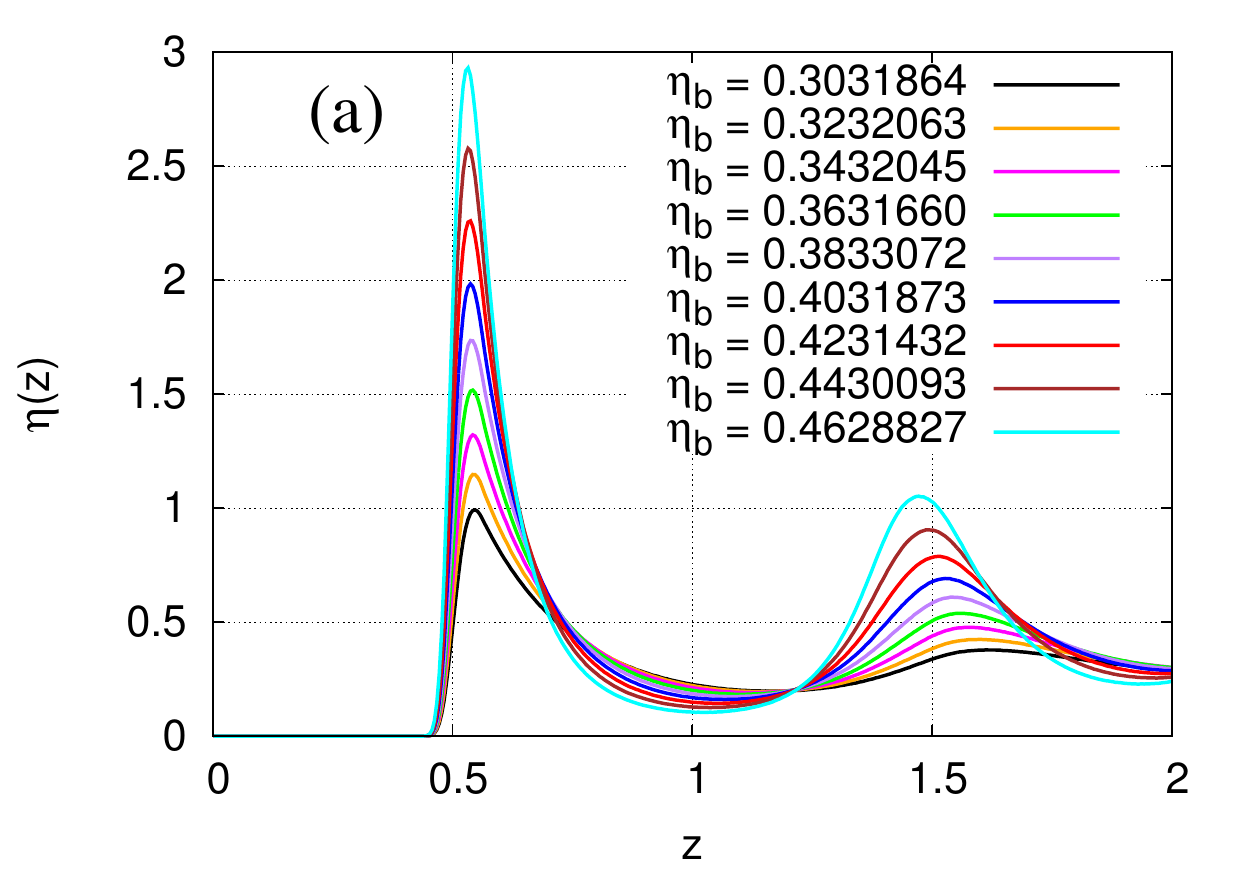} & \hspace*{0.0cm}&
\includegraphics[width=0.5\textwidth]{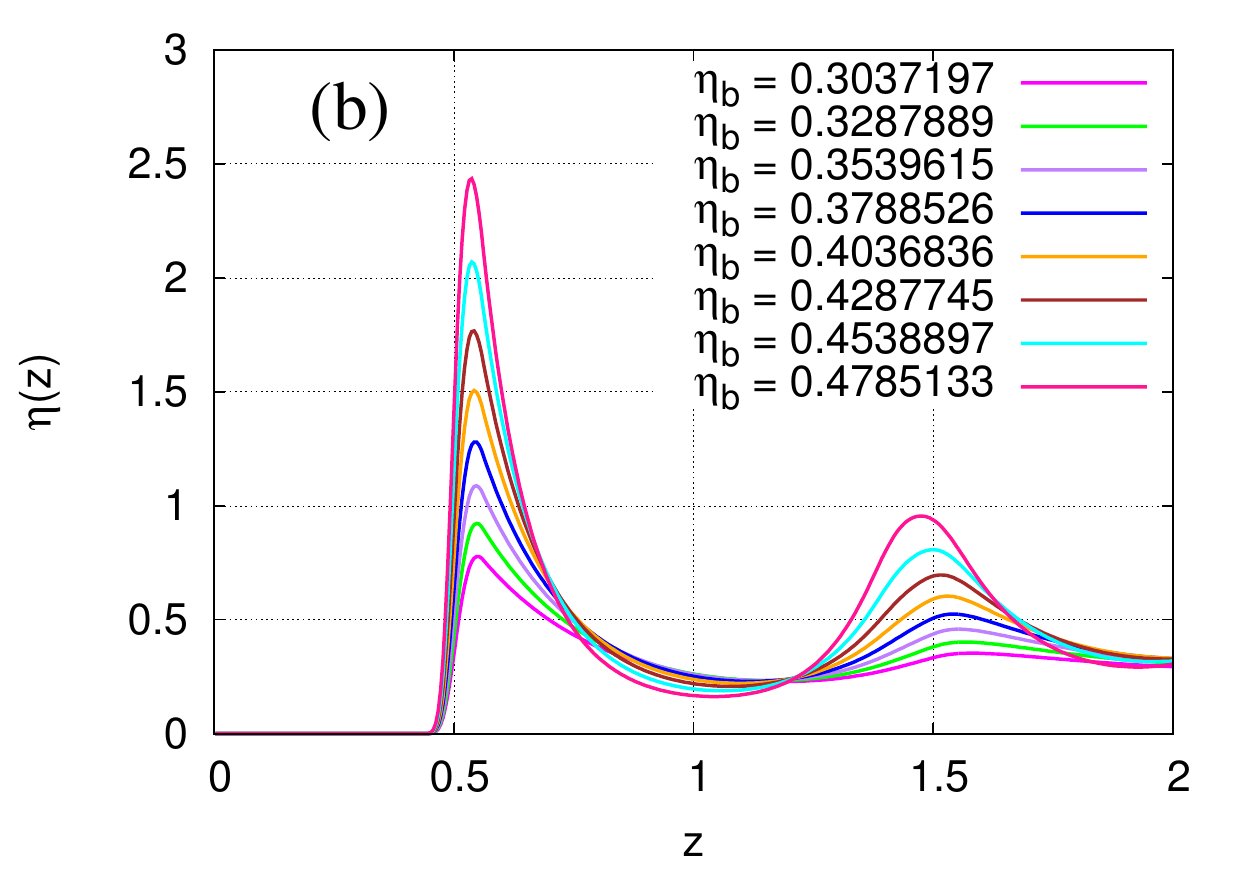}
\end{array}$
\caption{\label{fig4} Density profile $\rho(z)$ versus $z$ near WCA walls with $\epsilon=1$
at $z=0$, for the hard sphere model (a) and the AO model (b).}
\end{center}
\end{figure}

For the sampling of $P_{\rm nov}(\xi)$, a careful sampling of the probability that pairs of particles
occur at a distance $\sigma + \Delta r$ from each other is required, for very small values of $\Delta r$ is required.
Of course, the number of particles $n(\Delta r)$ that occur at a distance in between $\sigma$
and $\sigma + \Delta r$ goes to zero as $\Delta r \rightarrow 0$. The constants $b_N$ and $b_T$
in Eqs.~(\ref{eq15}),~(\ref{eq16}), as well as the analogous constant for the bulk pressure in a
system with periodic boundary conditions, depend very sensitively on the magnitude of $\xi$ that
is considered in the virtual volume changes considered in Eq.~(\ref{eq14}), or, equivalently,
the range of $\Delta r$ that is used to fit the derivative of $dn (\Delta r)/d(\Delta r)$ as
$\Delta r \rightarrow 0$. This problem is illustrated in Fig.~\ref{fig3} for several cases
of $\eta_b$. Note that the range of $\xi$ from $0< \xi < 5\cdot 10^{-4}$ must be probed to allow
an accurate linear fit from which the limiting value for $\xi \rightarrow 0$ can be extracted.
If one fits a larger range of $\xi$, systematic errors do arise, particularly for larger
values of the packing fraction. Note also that the relative range of pressure variation
with $\xi$ increases with increasing packing fraction. Thus, it is rather easy to find
a correct result for rather small values of the bulk packing fraction $\eta_b$, but
much more difficult when $\eta_b$ approaches the region where the onset of crystallization occurs.
This problem is not specific for the AO model, but occurs for the pure hard sphere system
as well. In fact, already an examination of the local density near the walls 
(Fig.~\ref{fig4}) shows that in the fluid phase both models are still very similar
(only the solid phases at the liquid to solid transition have rather different packing
fractions \cite{23}, and the interfacial stiffness between the coexisting phases
also differ appreciably \cite{23}). In the AO model, the first peak of the layering structure
near the WCA wall is slightly higher and slightly sharper as in the pure hard sphere case,
but otherwise the differences are rather minor.

\begin{figure}
\begin{center}$
\begin{array}{ccc}
\includegraphics[width=0.5\textwidth]{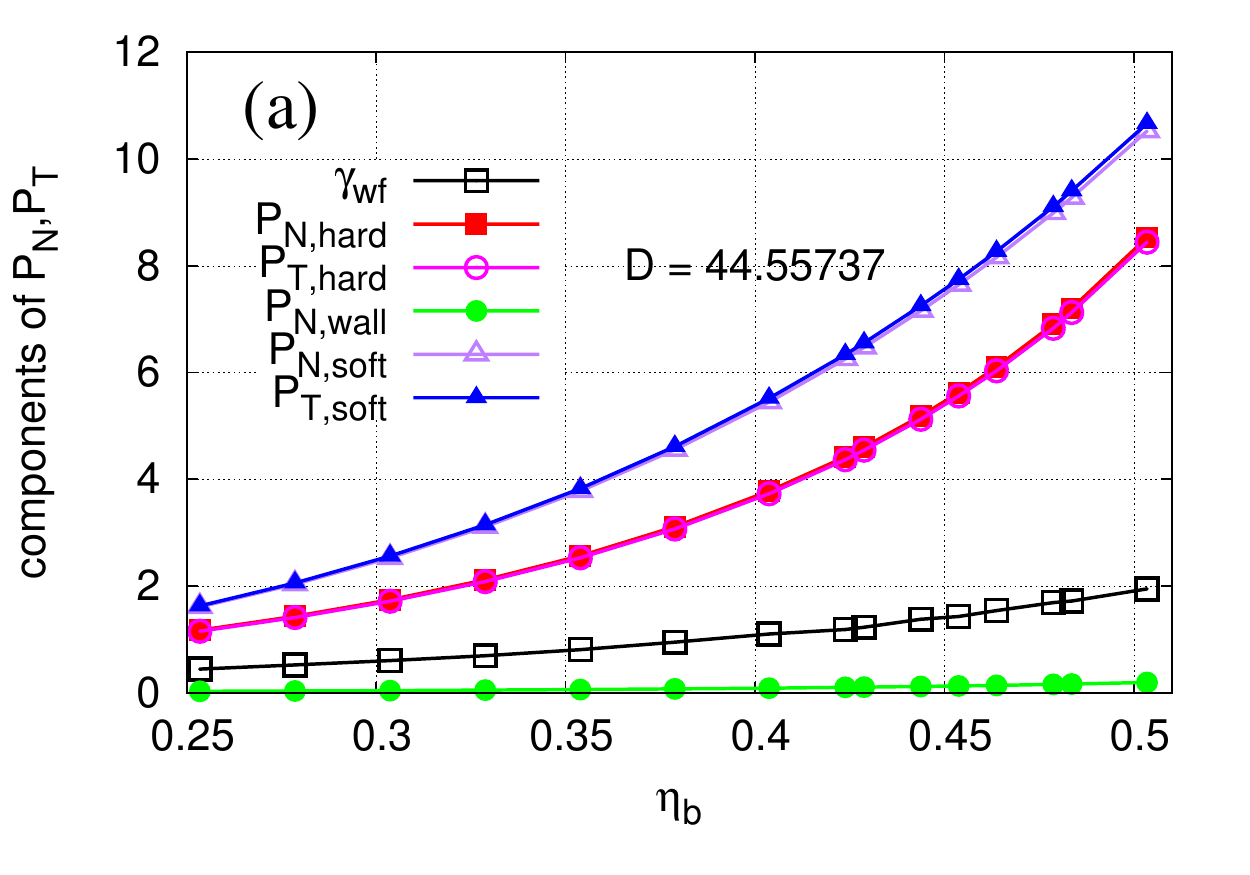} & \hspace*{0.0cm}&
\includegraphics[width=0.5\textwidth]{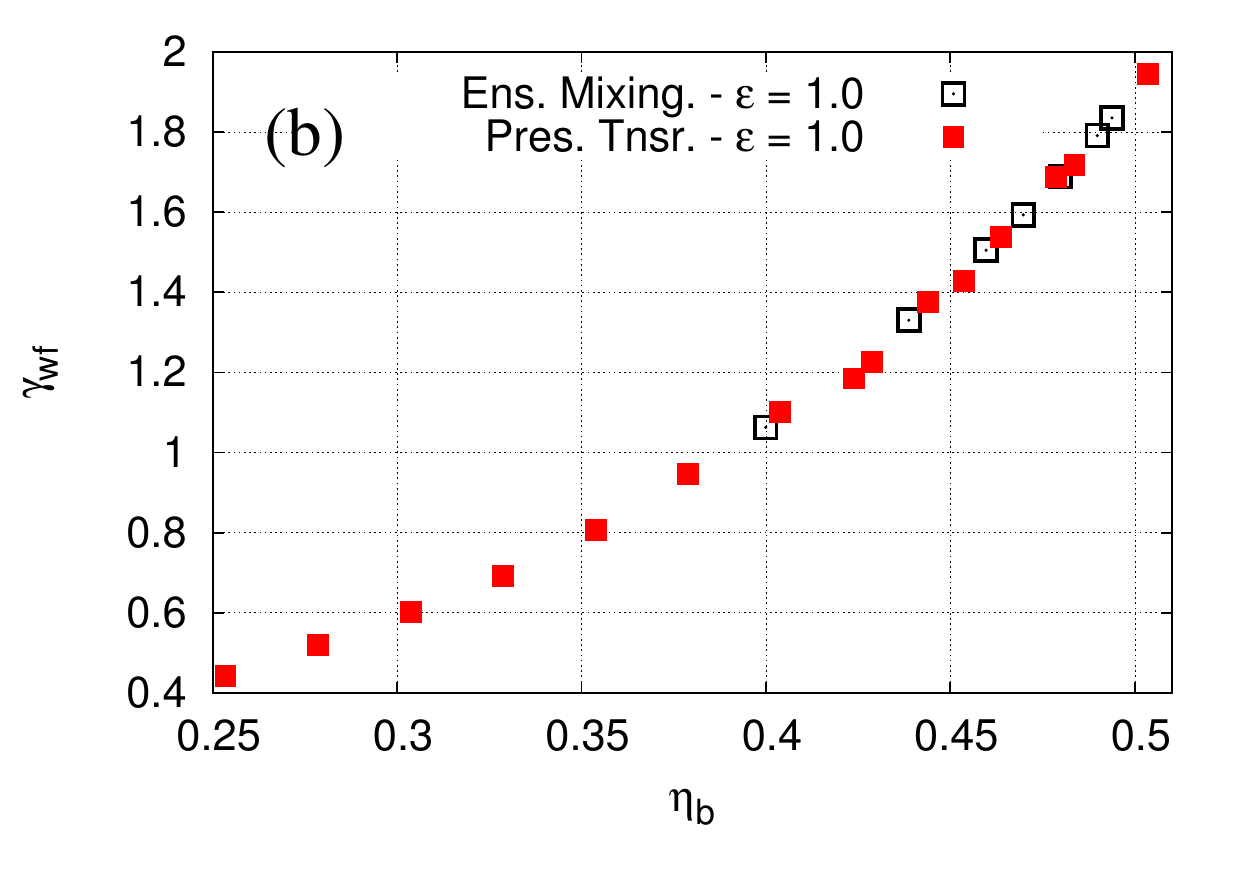}
\end{array}$
\caption{\label{fig5} (a) Various contributions to the normal pressure $P_N$ ($P_{N, {\rm wall}}$, $P_{N, {\rm int}}$)
and the tangential pressure $(P_T)$ for the AO model plotted versus $\eta_b$, choosing $D=44.55737$ and
two WCA walls with $\varepsilon=1.0$. (b) Wall tension $\gamma_{wf}$ plotted vs. $\eta_b$. Open
squares are due the ensemble mixing method, full squares result from the application of Eq.~(\ref{eq2}).}
\end{center}
\end{figure}

Fig.~\ref{fig5} then presents the various contributions to the pressure as well as the wall
tension $\gamma_{wf}$ as a function of bulk packing fraction. Of course, for the application
of Eq.~(\ref{eq2}) the individual terms $P_N$, $P_T$ need to be obtained with an accuracy
much better than one part in a thousand, if for large $D$ a meaningful accuracy of $\gamma_{wf}$
is desired: From Fig.~\ref{fig5}(a), we recognize that near the crystallization transition
(which occurs for \cite{23} $\eta_p=0.494)$ the pressure in the bulk is of order
10$^1$, while $\gamma_{wf}$ is of order 10$^o$. With $D/2$ of order 20, Eq.~(\ref{eq2})
implies $P_N-P_T$ is of order $1/20$, and hence $(P_N-P_T)/P_N$ is of order 1/200. This
consideration already explains why (on the scale of Fig.~\ref{fig5}a) the difference between
$P_N$ and $P_T$ is invisible, and getting $\gamma_{wf}$ accurate to a few percent hence is a
computational tour de force. This consideration explains readily why already for the
case of hard spheres it has happened that different methods for the estimation
of $\gamma_{wf}$ have yielded results that slightly different from each other (see Ref.~
\cite{13} for a recent account on the literature of this problem).

\section{The ensemble mixing method}

In Ref. \cite{13} we have already presented very briefly the ensemble mixing method
where one gradually transforms from a system without walls in the (NVT) ensemble
(with periodic boundary conditions in all three space directions) to a system with two
parallel walls at distance $D$ from each other (and periodic boundary conditions only in the $x,y$
directions parallel to the walls). Describing the system without walls by a Hamiltonian
$\mathcal{H}_1 (\vec{X})$, $\vec{X}$ denoting a point in configuration space, 
and the system with walls by $\mathcal{H}_2(\vec{X})$, the idea (originally proposed by Heni and L\"owen \cite{19})
is to construct a smooth path along which one can carry out a thermodynamic integration,
that yields the free energy difference between the two systems, from which hence $\gamma_{wf}$
can be inferred. Heni and L\"owen \cite{19} applied this idea to the fluid of
hard spheres, and also the recent extension due to Deb et al. \cite{13} deals with the
hard sphere fluid only. In the present section, we demonstrate that this method can equally
well be applied to the AO model.

Since both systems $\mathcal{H}_1(\vec{X})$ and $\mathcal{H}_2 (\vec{X})$ refer to the same
choice of particle number $N$ and volume $V$, i.e.~the same phase space, one can think
of constructing an interpolation Hamiltonian $\mathcal{H}(\vec{X})$ as follows,

\begin{equation} \label{eq19}
\mathcal{H}(\vec{X})= (1- \kappa) \mathcal{H}_1 (\vec{X}) + \kappa \mathcal{H}_2(\vec{X})
\end{equation}

where the parameter $\kappa \in [0,1]$ is varied for calculating the free energy
difference between the systems $(1,2)$. In our simulations, $\kappa$ is discretized
in a set of discrete values $\{\kappa_i\}$, and the system is allowed to jump from
$\kappa_i$ to a neighboring value, $\kappa_{i+1}$ or $\kappa_{i-1}$. These moves
are accepted or rejected with a Metropolis step, and hence in order to ensure a large
enough acceptance probability, the number of intermediate systems $\{\kappa_i\}$ needs
to be chosen proportional to the volume $V$ of the system. We sample the relative
probability $P(i)$ to reside in state $i$ with a variant of Wang-Landau sampling
\cite{24,25}. Alternatively, other free-energy schemes like successive umbrella 
sampling can be applied\cite{26}. The free energy difference between the two states $i$
and $i + 1$ is given by $k_BT [\ln P(i) - \ln P (i + 1)]$. In this way one can
sample the free energy difference $\Delta F(D)$ for any given choice of $D$,
between a system with walls and the system of corresponding thickness and particle
number but without walls. The wall free energy then follows as

\begin{equation} \label{eq20}
\gamma_{wf}=\lim\limits_{D \rightarrow \infty} \Delta F (D) / (2 L^2k_BT) \quad.
\end{equation}

\begin{figure}
\begin{center}$
\begin{array}{ccc}
\includegraphics[width=0.46\textwidth]{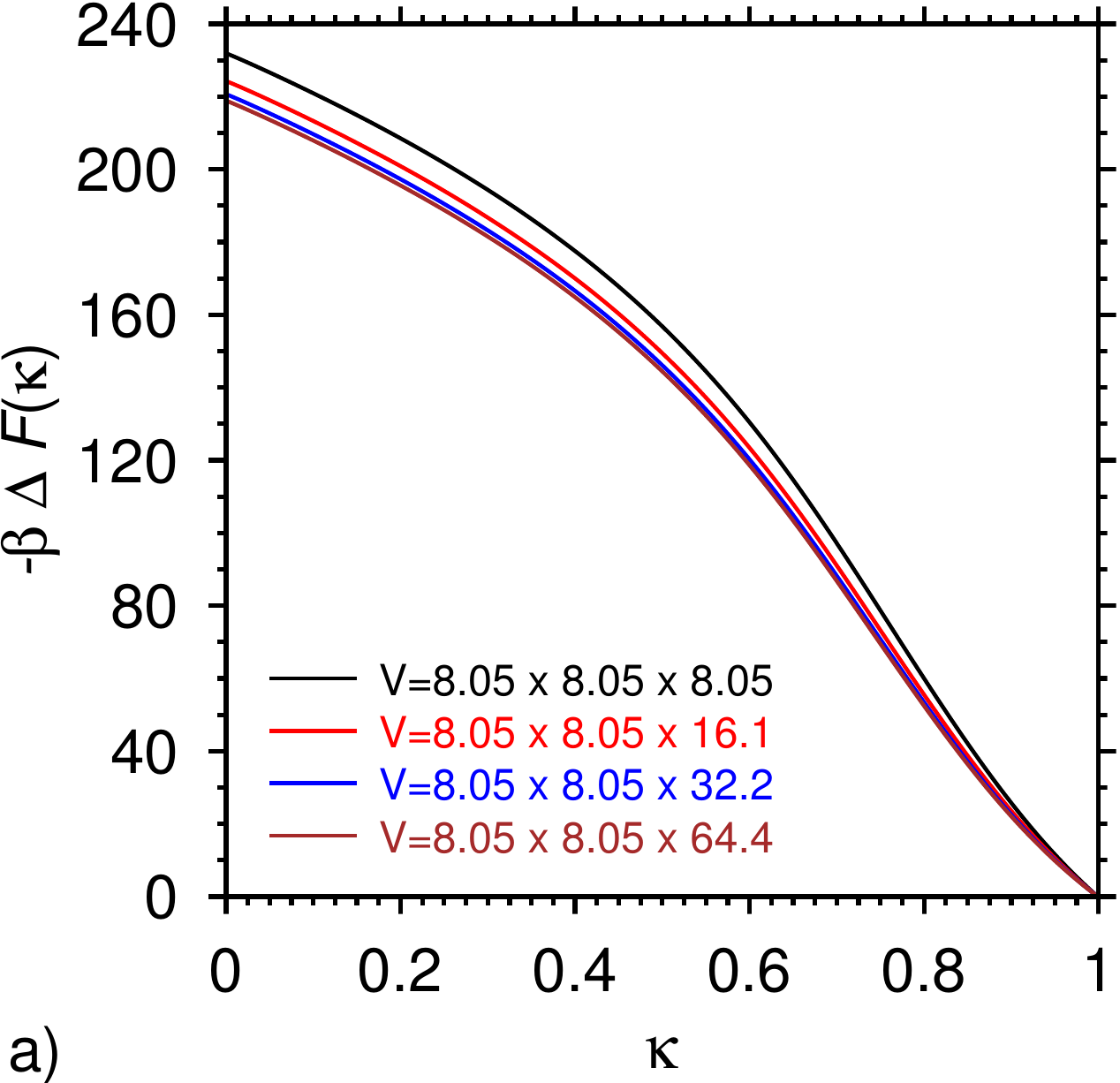} & \hspace*{0.05\textwidth}&
\includegraphics[width=0.46\textwidth]{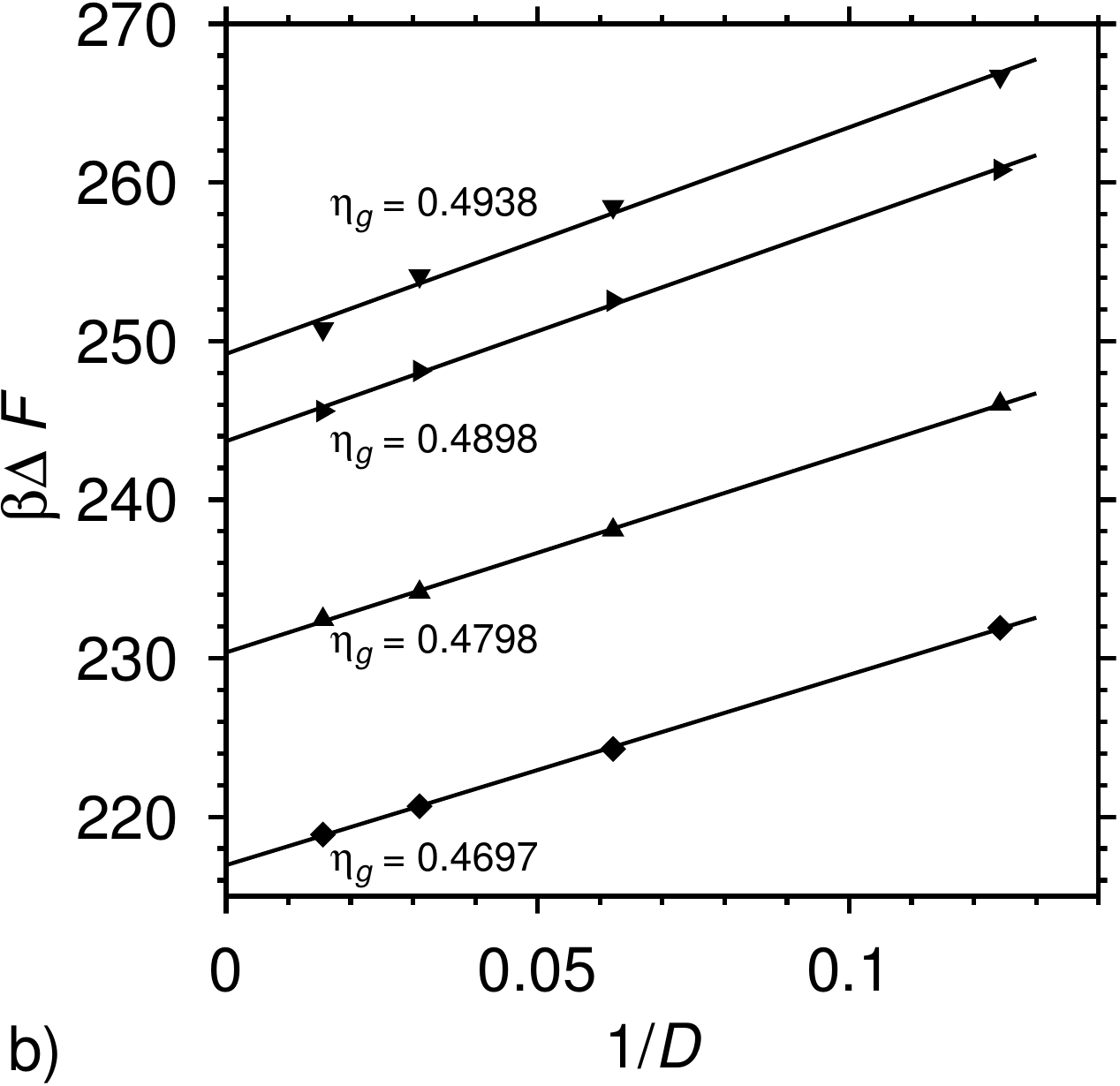}
\end{array}$
\caption{\label{fig6} Typical example showing how wall tensions $\gamma_{wf}$ are extracted using the
ensemble mixing method for the AO model. Part (a) shows the free energy difference $\Delta F(\kappa)/k_BT$ vs.
$\kappa$ for $\varepsilon=2$, $\eta_b=0.4697$. Part (b) shows the extrapolation to $1/ D \rightarrow 0$ for several $\eta_b$}.
\end{center}
\end{figure}

Note that due to the excess density $\rho_s$ at the walls (which is present in the systems
with walls, so that $\rho=N/V=\rho_b+ (2/D)\rho_s$), for any finite $D$ the bulk density of the
system with periodic boundary conditions (which is $\rho$) differs slightly but systematically
from the bulk density in the system with the walls (which is $\rho_b=\rho-(2/D)
\rho_s)$. This fact already suggests that the extrapolation in Eq.~(\ref{eq20}) should
be carried out in terms of a plot of $\Delta F(D)/(2L^2k_BT)$ linearly versus $1/D$
(Fig.~\ref{fig6}). In this way the data included already in Fig.~\ref{fig5}(b) have been
generated.

Of course, in this method the lateral linear dimension $(L=8.05 \sigma)$ is rather
small, and often even smaller choices for $L$ have been used (e.g., $L=5 \sigma)$
\cite{13}. However, no significant finite size effects associated with the use of
such small values of $L$ were detected. For the pressure tensor calculations,
somewhat larger choices for $L$ have been possible (near $L=12$). Again, the
good agreement between both methods to obtain $\gamma_{wf}$ for a significant
range of $\eta_b$ (Fig.~\ref{fig5}b) can be taken as an indication that neither
method suffers from significant systematic errors, in the range of the parameters
of interest for the present model.

\section{Conclusion}

In this paper, a comparative study of several methods to compute the excess free
energy of walls in fluids (the ``fluid-wall tension'') has been presented, considering
fluids where the particles exhibit a hard-core repulsion, and hence the standard
method based on the wall-induced anisotropy of the pressure tensor is not
straightforward to apply. We have described three methods: (i) In the Schmid-Schilling
method, the absolute free energy of the system is computed. When one does this for
systems in a film geometry with several choices for the distance $D$ between the walls
confining the film, one can infer the wall tension $\gamma_{wf}$ from the coefficient
of the term in the free energy that exhibits a $1/D$ variation. We have tested this
method so far for the hard sphere fluid only, but we did find good agreement with
other methods. (ii) The pressure tensor anisotropy method can in fact be generalized to systems,
with hard potentials, by a careful sampling of the probability $P_{\rm nov} (\xi)$ that
particles do not overlap when virtual changes of the linear dimensions by a
factor $1- \xi$ are performed. We have applied this method to the AO model of a
colloid-polymer mixture, where the (implicitly treated) polymers are responsible
for an effective attraction between the particles. The latter part of the potential
is treated via the virial formula for the pressure tensor, while the repulsive part
is treated with a sampling of $P_{\rm nov} (\xi)$, as mentioned above. We have shown
that by a careful consideration of accuracy issues this method does yield valid results,
but a substantial computational effort is needed. (iii) The third method is the ensemble
mixing method, where one considers systems with periodic boundary conditions and no walls
``mixed'' (in the sense of a combined Hamiltonian, Eq.~(\ref{eq19})) with a system with the
same particle number and linear dimensions, but bounded by walls. This method has been
used to test the other two methods, and our practical experience is that this last method
amounts to the relatively smallest computational effort, in comparison with the other
two methods. In the future work, we plan to use such methods to study the wetting
behavior of colloidal dispersions.\\

\underline{Acknowledgement}: This work was supported by Deutsche Forschungsgemeinschaft
(DFG) under grants No Bi 314/19 (D.D.), TR6/C4 (D.W.) and TR6/A5 (A.W.). We are
grateful to NIC J\"ulich for a generous grant of computer time at the JUROPA
supercomputer of the JSC. It is a pleasure to dedicate this paper to David P. Landau
on occasion of his 70$^{\rm th}$ birthday, and to thank him for his inspiring work
over many decades, which also has been instrumental for some of the developments
described here (in particular, the ensemble mixing method has profited from the
Wang-Landau algorithm).

\clearpage

\end{document}